\documentclass[aip,apl,reprint]{revtex4-1} 
\usepackage{amsmath}
\usepackage{amssymb}
\usepackage{bm}
\usepackage{color}
\usepackage{graphicx}
\usepackage[latin1]{inputenc}
\usepackage{times}
\usepackage{natbib}
\makeatletter

\begin{document}


\title{Frequency modulation of spin torque oscillator pairs} 




\author{Ye.~Pogoryelov}
\email{yevgenp@kth.se}
\affiliation{Materials Physics, Royal Institute of Technology, Electrum 229, 164 40 Kista, Sweden}

\author{P.~K.~Muduli}
\affiliation{Physics Department, University of Gothenburg, 412 96 Gothenburg, Sweden}

\author{S.~Bonetti}
\affiliation{Materials Physics, Royal Institute of Technology, Electrum 229, 164 40 Kista, Sweden}

\author{E.~Iacocca}
\affiliation{Physics Department, University of Gothenburg, 412 96 Gothenburg, Sweden}

\author{Fred~Mancoff}
\affiliation{Everspin Technologies, Inc., 1300 N. Alma School Road, Chandler, Arizona 85224, USA}

\author{Johan~\r{A}kerman}
\affiliation{Materials Physics, Royal Institute of Technology, Electrum 229, 164 40 Kista, Sweden}

\affiliation{Physics Department, University of Gothenburg, 412 96 Gothenburg, Sweden}


\date{\today}

\begin{abstract}
The current controlled modulation of nano-contact based spin torque oscillator (STO) pairs is studied in both the synchronized and non-synchronized states. The synchronized state shows a well behaved modulation and demonstrates robust mutual locking even under strong modulation. The power distribution of the modulation sidebands can be quantitatively described by assuming a \emph{single} oscillator model. However, in the non-synchronized state, the modulation sidebands are not well described by the model, indicating interactions between the two individual nano-contact STOs. These findings are promising for potential applications requiring the modulation of large synchronized STO arrays.
\end{abstract}

\pacs{72.25.Ba, 75.75.+a, 75.20.$-$g, 85.75.$-$d}

\maketitle 


Spin Torque Oscillators (STOs) are attracting a rapidly growing interest due to their potential use in future microwave and memory applications.~\cite{Braganca2010, Katine2008JMMM, Silva2008JMMM} The principle of the STO operation is based on the transfer of spin angular momentum from a spin polarized current to the local magnetization of a thin magnetic layer.~\cite{Slonczewski1996,Berger1996} The effect is typically achieved in a nanoscale device ($\sim 100$~nm in lateral size) in which a large current density ($\sim10^{8}$ A/cm$^{2}$) can generate a precession of the magnetization. This precession is then detected by the oscillation of the giant magnetoresistance (GMR) or tunneling magnetoresistance (TMR) of the device.

However, the low output power of the STOs still limits their use in actual applications. A promising way to increase the STO output power is the synchronization of separate nano-contact STOs that share a common magnetic layer.~\cite{Kaka2005,Mancoff2005}
Spin waves generated beneath each nano-contact propagate in the common layer providing a coupling mechanism that can phase-lock the oscillators and coherently sum their individual amplitudes.~\cite{pufallmr2006,Slavin2006,Rezende2007prl,chen2009prb}
Furthermore, it is imperative to demonstrate that the signal from several synchronized STOs can be modulated without perturbing, or even unlocking, the synchronized state. While the modulation of single nano-contact based STOs has been both studied theoretically~\cite{SlavinTutorial,consolo2010ieee} and demonstrated experimentally~\cite{Pufall2005APL,muduli2010prb} and the literature on the theoretical aspects of STO synchronization has been growing steadily,~\cite{Slavin2005aPRB,Slavin2006,SlavinTutorial,Rezende2007prl,chen2009prb,grollierj2006,Tiberkevich2009apl,Bonin2009EPJB,Sano2007jjap,Georges2008,Persson2007,Zhou2007,Zhou2008,Zhou2009ieee,zhou2009apl} neither experimental nor theoretical studies of modulated synchronized STOs have been presented.

In this work we study the current controlled modulation of nano-contact based STO pairs in their synchronized and non-synchronized states. In the non-synchronized state the individual STOs are modulated independently and exhibit separate sidebands that are not well modeled using prior modulation theories. We ascribe this discrepancy to interactions between the STOs. However, the modulation of the synchronized state is well behaved and can be quantitatively explained assuming a \emph{single} STO model. Additionally, the robustness of the synchronization, even under strong modulation, is encouraging for future applications based on the modulation of large STO arrays.

\begin{figure}[t!]
\centering
\includegraphics[width=0.35\textwidth]{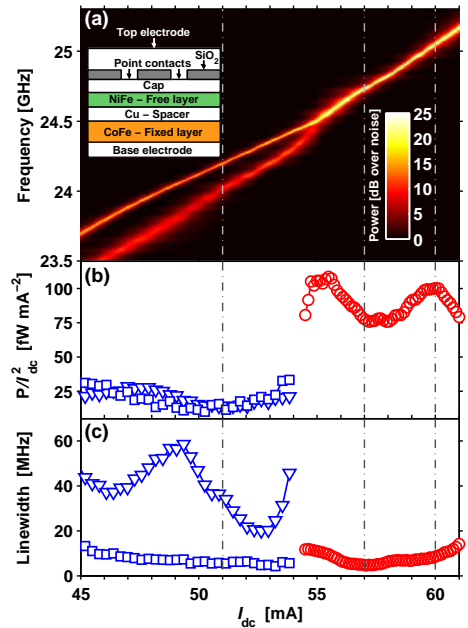}
\caption{(Color online) Current dependence of the free-running nano-contact STO pair: (a) Map of the peak power versus frequency and applied current bias $I_{dc}$. Peak power is expressed in dB over the noise floor. Inset shows a schematic cross-section of the two nano-contacts on a GMR spin-valve mesa. (b) Normalized integrated power, $P/I^{2}_{dc}$ and (c) linewidth versus $I_{dc}$, where the data below the synchronization transition ($I_{dc} <$ 54 mA) is given by triangles and squares for the lower and higher frequency peaks respectively, and within the synchronized region ($I_{dc} >$ 54 mA) by circles. Vertical dashed lines indicate the three operating points, $I_{dc}$ = 51, 57 and 60 mA, used to compare modulation in the non-synchronized and synchronized states.}
\label{fig:fig1}
\end{figure}

The particular device studied in this work has two nano-contacts, nominally 80 nm in diameter, with 400 nm center-to-center separation (see schematic cross-section in the inset in Fig.~\ref{fig:fig1}(a)). Such nanocontact based STO pairs have been described in detail in Ref.~\onlinecite{Mancoff2005}.

The low frequency (100 MHz) modulating current is injected from an RF source to the STO via a circulator. The \textit{dc} bias current is fed to the device by a precision current source (Keithley 6221) through a 0-40~GHz bias tee connected in parallel with the transmission line. The signal is then amplified using a broadband 16-40~GHz, +22~dB microwave amplifier, and finally detected by a spectrum analyzer (Rohde \& Schwarz FSU46).
For more details see Ref.~\onlinecite{muduli2010prb}. All measurements were performed in a magnetic field of 10~kOe applied at an angle of 66$^{\circ}$ to the film plane to both maximize the output power~\cite{bonetti2009apl} and ensure that a single, propagating, spin wave mode is excited.~\cite{Slonczewski1999, Slavin2005PRL, Bonetti2010prl}

Fig.~\ref{fig:fig1}(a) shows the power spectral density (PSD) of the STO signal as a function of drive current $I_{dc}$. A clear transition from a non-synchronized two-signal regime to a single synchronized state can be observed at about $I_{dc}=54$~mA, where the total normalized power $P/I^{2}_{dc}$ approximately doubles from 43~nW/A$^2$ to 91.1~nW/A$^2$ [Fig.~\ref{fig:fig1}(b)], and the average linewidth drops significantly [Fig.~\ref{fig:fig1}(c)]. The doubling of the total output power indicates that the amplitudes from the individual nano-contacts add coherently, and in phase, and the decrease in average linewidth is consistent with an increase in the precessional mode volume, rendering thermal fluctuations less effective in perturbing the precession orbit.~\cite{Mancoff2005,Kaka2005,Russek2005PRB,Silva2010ieee}

For the modulation experiment below, we chose three operating points well within the non-synchronized ($I_{dc}=51$~mA) and synchronized ($I_{dc}=57$~and 60~mA) regimes, respectively, indicated by dashed lines in Fig.~\ref{fig:fig1}(a). Operating points within the synchronized regime were selected to match local minimum of linewidth ($I_{dc}=57$~mA) and local maximum of integrated power ($I_{dc}=60$~mA). Modulation was studied up to a maximum modulation current of $I_{m}=3$~mA to ensure that the STO pair stays within a given state at all times during modulation for operating points at $I_{dc}=$~51 and 57~mA. For the operating point at $I_{dc}=60$~mA modulation was studied up to $I_{m}=4.5$~mA.

\begin{figure}[t!]
\centering
\includegraphics[width=0.35\textwidth]{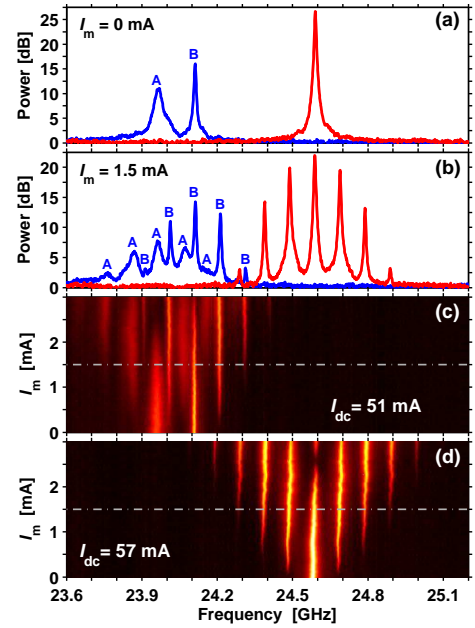}
\caption{(Color online) (a) PSD of the free-running STO for $I_{dc}=$~51 mA (A, B) and 57 mA; (b) Same spectra when a modulating current ($I_{m}=1.5$~mA, $f_{m}=100$~MHz) is added. \textquotedblleft A\textquotedblright~and \textquotedblleft B\textquotedblright~denote the individual STO signals in the non-synchronized state. (c) and (d) show color maps of the PSD vs. $I_{m}$ in the non-synchronized and synchronized states, respectively. Color scale is identical to Fig.~\ref{fig:fig1}(a).}
\label{fig:fig2}
\end{figure}

Fig.~\ref{fig:fig2} shows the result of current modulation. The free-running STO spectra [Fig.~\ref{fig:fig2}(a)] develop a number of sidebands equally spaced around each carrier signal [Fig.~\ref{fig:fig2}(b)]. In the non-synchronized state, it is possible to identify all sidebands as belonging to either the low-frequency carrier (\textquotedblleft A\textquotedblright) or the high-frequency carrier (\textquotedblleft B\textquotedblright). In the synchronized state, the sidebands look remarkably similar to what one would expect from a single STO and there is no indication of any unlocking. This is further emphasized by plotting color maps of the resulting PSD as a function of $I_{m}$ [Fig.~\ref{fig:fig2}(c) and (d)]. In particular, the synchronized state shows very sharp and well formed sidebands at all investigated modulation currents.

\begin{figure}[t!]
\centering
\includegraphics[width=0.35\textwidth]{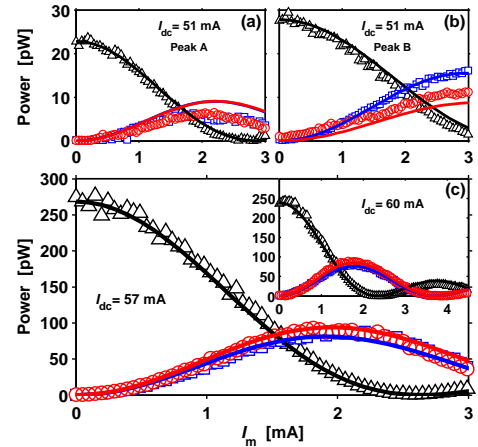}
\caption{(Color online) Integrated power of the carrier (triangles) and the first order upper (squares) and lower (circles) sidebands for (a) Peak A and (b) Peak B for $I_{dc}$ = 51~mA and (c) for $I_{dc}$ = 57~mA. Inset in (c) shows experimental and calculated values of integrated power for $I_{dc}$ = 60~mA. Solid lines represent the corresponding calculated integrated power from NFAM theory.}
\label{fig:fig3}
\end{figure}

Fig.~\ref{fig:fig3} shows the integrated output power of the carrier (triangles) and the first order upper (squares) and lower (circles) sidebands. Fig.~\ref{fig:fig3}(a) and (b) shows data for the non-synchronized regime where there is a substantial difference in the sidebands of peak \textquotedblleft B\textquotedblright, indicating rather strong non-linearities for this STO. In the synchronized regime, on the other hand, the first-order sideband amplitudes are found to be nearly equal, indicating a highly linear behavior of $f$($I_{dc}$). This possibly suggests that the non-linearities of the individual STOs can be averaged out once they synchronize. The amplitude modulation is also found to be negligible for the first-order sidebands. However, when analyzing higher order sidebands, for which amplitude modulation should have a stronger impact,~\cite{consolo2010ieee} we indeed observed a measurable difference (not shown), consistent with the amplitude nonlinearity around $I_{dc}=57$~mA [Fig.~\ref{fig:fig1}(b)].

Following the method developed in Ref.~\onlinecite{muduli2010prb} we now use the free-running STO characteristics [Fig.~\ref{fig:fig1}] to calculate the theoretically expected sidebands from a combined nonlinear frequency and amplitude modulation (NFAM) theory~\cite{muduli2010prb,consolo2010ieee} without any free parameters. In the synchronized state [Fig.~\ref{fig:fig3}(c)], we essentially find perfect agreement between experiment and calculations, which further corroborates that two synchronized STOs behave as a single STO under modulation. In the non-synchronized state, on the other hand, experiment and calculations show much less convincing agreement [Fig.~\ref{fig:fig3}(a) and (b)]. We ascribe this discrepancy to strong interaction between the two nano-contact STOs even in the non-synchronized state.
The interaction (and mutual phase locking) under modulation can be qualitatively investigated by following the procedure of Ref.~\onlinecite{SlavinTutorial} (section VI.B) for similar nano-contacts considering that the individual STOs are modulated (Eq.~35)
\begin{equation}
\label{eq:slavin}
    \frac{d\Psi}{dt} = 2\pi\Delta f - 2\mu\nu\Delta\Gamma(p)\cos(2\pi f_mt) - \Delta_2 \sin(\Psi),
\end{equation}
where $\Psi$ is the instantaneous phase difference between the oscillators, $\Delta f$ is the frequency mismatch, $\mu = I_m/I_{dc}$, $\nu$ is the dimensionless nonlinear frequency shift, $\Delta\Gamma(p)$ is the difference in the power dependent restoration rate, and $\Delta_2$ is the phase locking bandwidth. In the free-running case ($\mu=0$), Eq.~\ref{eq:slavin} reduces to Eq. 65(b) in Ref.~\onlinecite{SlavinTutorial}, and describes the regimes of mutual phase locking ($2\pi\Delta f<\Delta_2$), mutual frequency pulling ($2\pi\Delta f>\Delta_2$) and non-interacting STOs ($2\pi\Delta f\gg\Delta_2$). In the mutual frequency pulling regime, the solution of Eq.~\ref{eq:slavin}, although unbounded, has a characteristic \emph{beat frequency} which corresponds to the \emph{measured} $\Delta f$.~\cite{Pikovsky2001} Using the experimental data in Fig.~\ref{fig:fig2}(a), we can estimate this beat frequency to lie between $100 - 160$~MHz in the modulated region around $I_{dc}=51$~mA. As a consequence, even in the free-running case, the two STOs are intrinsically self-modulated, or intermodulated, by their mutual interaction. Since the NFAM calculation as its starting point assumes an entirely non-modulated free-running state, it fails to accurately predict the modulated sideband powers in a situation where intrinsic mutual modulation both affects the free-running state, and through Eq.~\ref{eq:slavin} continues to affect the modulated state in a highly non-trivial fashion.

The free-running synchronized state, on the other hand, is characterized by a constant $\Psi=\Psi_0$, without any mutual modulation. Linearizing Eq.~\ref{eq:slavin} around $\Psi_0$, we get for the modulated state,
\begin{eqnarray}
\label{eq:solution}
    \Psi \approx \Psi_0 + \frac{2\mu\nu\Delta\Gamma(p)}{\Delta_2^2+(2\pi f_m)^2}(2\pi f_m\sin(2\pi f_mt)\nonumber \\
    +\Delta_2\cos(2\pi f_mt)),
\end{eqnarray}
where the second term introduces a small time-varying phase difference between the synchronized STOs. As long as this term is small, the extrinsic modulation has no impact on the synchronized state, and NFAM for a single STO is applicable. However, Eq.~\ref{eq:solution} suggests that there should exist a modulation strength, $\mu$, for which the driven variation in $\Psi$, ($\Delta\Psi$) becomes larger than its thermal fluctuation and deviations from NFAM theory could potentially be observed. A further increase of $\mu$ may even unlock the oscillators when $\Delta\Psi > \pi$ in which case phase slips occur. Since we observe good agreement with NFAM theory, we argue that under a modulation of $I_m = 4.5$~mA at $f_m = 100$~MHz, $\Delta\Psi$ is negligible.

In conclusion, we have studied frequency modulation of nano-contact based spin torque oscillator pairs. In the non-synchronized state, the individual STOs are modulated separately and exhibit a complex sideband behavior, which we ascribe to interactions between the two nano-contact STOs not considered by NFAM. At the same time the synchronized state demonstrates outstanding stability under modulation. In addition, the modulated response of synchronized STOs shows remarkable agreement with NFAM theory derived for \emph{single} STOs. Synchronized STOs can consequently both be modulated and modeled in a straightforward way as they behave as ordinary single RF oscillators at the investigated modulation frequency. We believe these results are important for the continued development of communication and signal processing applications based on spin torque oscillators.

Support from the Swedish Foundation for Strategic Research (SSF), the Swedish Research Council (VR), the G\"{o}ran Gustafsson Foundation and the Knut and Alice Wallenberg Foundation is gratefully acknowledged. Johan~\r{A}kerman is a Royal Swedish Academy of Sciences Research Fellow supported by a grant from the Knut and Alice Wallenberg Foundation. We thank
Randy K. Dumas for critical reading of the manuscript.


%


\begin{thebibliography}{33}%
\makeatletter
\providecommand \@ifxundefined [1]{%
 \@ifx{#1\undefined}
}%
\providecommand \@ifnum [1]{%
 \ifnum #1\expandafter \@firstoftwo
 \else \expandafter \@secondoftwo
 \fi
}%
\providecommand \@ifx [1]{%
 \ifx #1\expandafter \@firstoftwo
 \else \expandafter \@secondoftwo
 \fi
}%
\providecommand \natexlab [1]{#1}%
\providecommand \enquote  [1]{``#1''}%
\providecommand \bibnamefont  [1]{#1}%
\providecommand \bibfnamefont [1]{#1}%
\providecommand \citenamefont [1]{#1}%
\providecommand \href@noop [0]{\@secondoftwo}%
\providecommand \href [0]{\begingroup \@sanitize@url \@href}%
\providecommand \@href[1]{\@@startlink{#1}\@@href}%
\providecommand \@@href[1]{\endgroup#1\@@endlink}%
\providecommand \@sanitize@url [0]{\catcode `\\12\catcode `\$12\catcode
  `\&12\catcode `\#12\catcode `\^12\catcode `\_12\catcode `\%12\relax}%
\providecommand \@@startlink[1]{}%
\providecommand \@@endlink[0]{}%
\providecommand \url  [0]{\begingroup\@sanitize@url \@url }%
\providecommand \@url [1]{\endgroup\@href {#1}{\urlprefix }}%
\providecommand \urlprefix  [0]{URL }%
\providecommand \Eprint [0]{\href }%
\@ifxundefined \urlstyle {%
  \providecommand \doi  [0]{\begingroup \@sanitize@url \@doi}%
  \providecommand \@doi [1]{\endgroup \@@startlink {\doibase
  #1}doi:\discretionary {}{}{}#1\@@endlink }%
}{%
  \providecommand \doi  [0]{doi:\discretionary{}{}{}\begingroup
  \urlstyle{rm}\Url }%
}%
\providecommand \doibase [0]{http://dx.doi.org/}%
\providecommand \Doi [0]{\begingroup \@sanitize@url \@Doi }%
\providecommand \@Doi  [1]{\endgroup\@@startlink{\doibase#1}\@@Doi}%
\providecommand \@@Doi [1]{#1\@@endlink}%
\providecommand \selectlanguage [0]{\@gobble}%
\providecommand \bibinfo  [0]{\@secondoftwo}%
\providecommand \bibfield  [0]{\@secondoftwo}%
\providecommand \translation [1]{[#1]}%
\providecommand \BibitemOpen [0]{}%
\providecommand \bibitemStop [0]{}%
\providecommand \bibitemNoStop [0]{.\EOS\space}%
\providecommand \EOS [0]{\spacefactor3000\relax}%
\providecommand \BibitemShut  [1]{\csname bibitem#1\endcsname}%
\bibitem [{\citenamefont {Braganca}\ \emph {et~al.}(2010)\citenamefont
  {Braganca}, \citenamefont {Gurney}, \citenamefont {Wilson}, \citenamefont
  {Katine}, \citenamefont {Maat},\ and\ \citenamefont
  {Childress}}]{Braganca2010}%
  \BibitemOpen
  \bibfield  {author} {\bibinfo {author} {\bibfnamefont {P.~M.}\ \bibnamefont
  {Braganca}}, \bibinfo {author} {\bibfnamefont {B.~A.}\ \bibnamefont
  {Gurney}}, \bibinfo {author} {\bibfnamefont {B.~A.}\ \bibnamefont {Wilson}},
  \bibinfo {author} {\bibfnamefont {J.~A.}\ \bibnamefont {Katine}}, \bibinfo
  {author} {\bibfnamefont {S.}~\bibnamefont {Maat}}, \ and\ \bibinfo {author}
  {\bibfnamefont {J.~R.}\ \bibnamefont {Childress}},\ }
  {\bibfield  {journal}
  {\bibinfo  {journal} {Nanotechnology},\ }\textbf {\bibinfo {volume} {21}},\
  \bibinfo {pages} {235202} (\bibinfo {year} {2010})}\BibitemShut {NoStop}%
\bibitem [{\citenamefont {Katine}\ and\ \citenamefont
  {Fullerton}(2008)}]{Katine2008JMMM}%
  \BibitemOpen
  \bibfield  {author} {\bibinfo {author} {\bibfnamefont {J.~A.}\ \bibnamefont
  {Katine}}\ and\ \bibinfo {author} {\bibfnamefont {E.~E.}\ \bibnamefont
  {Fullerton}},\ }
  {\bibfield  {journal} {\bibinfo  {journal} {J. Magn. Magn. Mater.},\ }\textbf
  {\bibinfo {volume} {320}},\ \bibinfo {pages} {1217} (\bibinfo {year}
  {2008})}\BibitemShut {NoStop}%
\bibitem [{\citenamefont {Silva}\ and\ \citenamefont
  {Rippard}(2008)}]{Silva2008JMMM}%
  \BibitemOpen
  \bibfield  {author} {\bibinfo {author} {\bibfnamefont {T.~J.}\ \bibnamefont
  {Silva}}\ and\ \bibinfo {author} {\bibfnamefont {W.~H.}\ \bibnamefont
  {Rippard}},\ }
  {\bibfield  {journal} {\bibinfo  {journal} {J.
  Magn. Magn. Mater.},\ }\textbf {\bibinfo {volume} {320}},\ \bibinfo {pages}
  {1260} (\bibinfo {year} {2008})}\BibitemShut {NoStop}%
\bibitem [{\citenamefont {Slonczewski}(1996)}]{Slonczewski1996}%
  \BibitemOpen
  \bibfield  {author} {\bibinfo {author} {\bibfnamefont {J.~C.}\ \bibnamefont
  {Slonczewski}},\ }
  {\bibfield  {journal} {\bibinfo  {journal} {J.
  Magn. Magn. Mater.},\ }\textbf {\bibinfo {volume} {159}},\ \bibinfo {pages}
  {L1} (\bibinfo {year} {1996})}\BibitemShut {NoStop}%
\bibitem [{\citenamefont {Berger}(1996)}]{Berger1996}%
  \BibitemOpen
  \bibfield  {author} {\bibinfo {author} {\bibfnamefont {L.}~\bibnamefont
  {Berger}},\ }
  {\bibfield  {journal} {\bibinfo  {journal} {Phys.
  Rev. B},\ }\textbf {\bibinfo {volume} {54}},\ \bibinfo {pages} {9353}
  (\bibinfo {year} {1996})}\BibitemShut {NoStop}%
\bibitem [{\citenamefont {Kaka}\ \emph {et~al.}(2005)\citenamefont {Kaka},
  \citenamefont {Pufall}, \citenamefont {Rippard}, \citenamefont {Silva},
  \citenamefont {Russek},\ and\ \citenamefont {Katine}}]{Kaka2005}%
  \BibitemOpen
  \bibfield  {author} {\bibinfo {author} {\bibfnamefont {S.}~\bibnamefont
  {Kaka}}, \bibinfo {author} {\bibfnamefont {M.~R.}\ \bibnamefont {Pufall}},
  \bibinfo {author} {\bibfnamefont {W.~H.}\ \bibnamefont {Rippard}}, \bibinfo
  {author} {\bibfnamefont {T.~J.}\ \bibnamefont {Silva}}, \bibinfo {author}
  {\bibfnamefont {S.~E.}\ \bibnamefont {Russek}}, \ and\ \bibinfo {author}
  {\bibfnamefont {J.~A.}\ \bibnamefont {Katine}},\ }
  {\bibfield  {journal}
  {\bibinfo  {journal} {Nature},\ }\textbf {\bibinfo {volume} {437}},\ \bibinfo
  {pages} {389} (\bibinfo {year} {2005})}\BibitemShut {NoStop}%
\bibitem [{\citenamefont {Mancoff}\ \emph {et~al.}(2005)\citenamefont
  {Mancoff}, \citenamefont {Rizzo}, \citenamefont {Engel},\ and\ \citenamefont
  {Tehrani}}]{Mancoff2005}%
  \BibitemOpen
  \bibfield  {author} {\bibinfo {author} {\bibfnamefont {F.~B.}\ \bibnamefont
  {Mancoff}}, \bibinfo {author} {\bibfnamefont {N.~D.}\ \bibnamefont {Rizzo}},
  \bibinfo {author} {\bibfnamefont {B.~N.}\ \bibnamefont {Engel}}, \ and\
  \bibinfo {author} {\bibfnamefont {S.}~\bibnamefont {Tehrani}},\ }
  {\bibfield  {journal}
  {\bibinfo  {journal} {Nature},\ }\textbf {\bibinfo {volume} {437}},\ \bibinfo
  {pages} {393} (\bibinfo {year} {2005})}\BibitemShut {NoStop}%
\bibitem [{\citenamefont {Pufall}\ \emph {et~al.}(2006)\citenamefont {Pufall},
  \citenamefont {Rippard}, \citenamefont {Russek}, \citenamefont {Kaka},\ and\
  \citenamefont {Katine}}]{pufallmr2006}%
  \BibitemOpen
  \bibfield  {author} {\bibinfo {author} {\bibfnamefont {M.~R.}\ \bibnamefont
  {Pufall}}, \bibinfo {author} {\bibfnamefont {W.~H.}\ \bibnamefont {Rippard}},
  \bibinfo {author} {\bibfnamefont {S.~E.}\ \bibnamefont {Russek}}, \bibinfo
  {author} {\bibfnamefont {S.}~\bibnamefont {Kaka}}, \ and\ \bibinfo {author}
  {\bibfnamefont {J.~A.}\ \bibnamefont {Katine}},\ }
  {\bibfield  {journal} {\bibinfo  {journal}
  {Phys. Rev. Lett.},\ }\textbf {\bibinfo {volume} {97}},\ \bibinfo {pages}
  {087206} (\bibinfo {year} {2006})}\BibitemShut {NoStop}%
\bibitem [{\citenamefont {Slavin}\ and\ \citenamefont
  {Tiberkevich}(2006)}]{Slavin2006}%
  \BibitemOpen
  \bibfield  {author} {\bibinfo {author} {\bibfnamefont {A.~N.}\ \bibnamefont
  {Slavin}}\ and\ \bibinfo {author} {\bibfnamefont {V.~S.}\ \bibnamefont
  {Tiberkevich}},\ }
  {\bibfield  {journal} {\bibinfo  {journal}
  {Phys. Rev. B},\ }\textbf {\bibinfo {volume} {74}},\ \bibinfo {pages}
  {104401} (\bibinfo {year} {2006})}\BibitemShut {NoStop}%
\bibitem [{\citenamefont {Rezende}\ \emph {et~al.}(2007)\citenamefont
  {Rezende}, \citenamefont {de~Aguiar}, \citenamefont
  {Rodr\'\i{}guez-Su\'arez},\ and\ \citenamefont {Azevedo}}]{Rezende2007prl}%
  \BibitemOpen
  \bibfield  {author} {\bibinfo {author} {\bibfnamefont {S.~M.}\ \bibnamefont
  {Rezende}}, \bibinfo {author} {\bibfnamefont {F.~M.}\ \bibnamefont
  {de~Aguiar}}, \bibinfo {author} {\bibfnamefont {R.~L.}\ \bibnamefont
  {Rodr\'\i{}guez-Su\'arez}}, \ and\ \bibinfo {author} {\bibfnamefont
  {A.}~\bibnamefont {Azevedo}},\ }
  {\bibfield
  {journal} {\bibinfo  {journal} {Phys. Rev. Lett.},\ }\textbf {\bibinfo
  {volume} {98}},\ \bibinfo {pages} {087202} (\bibinfo {year}
  {2007})}\BibitemShut {NoStop}%
\bibitem [{\citenamefont {Chen}\ and\ \citenamefont
  {Victora}(2009)}]{chen2009prb}%
  \BibitemOpen
  \bibfield  {author} {\bibinfo {author} {\bibfnamefont {X.}~\bibnamefont
  {Chen}}\ and\ \bibinfo {author} {\bibfnamefont {R.~H.}\ \bibnamefont
  {Victora}},\ }
  {\bibfield  {journal} {\bibinfo  {journal}
  {Phys. Rev. B},\ }\textbf {\bibinfo {volume} {79}},\ \bibinfo {pages}
  {180402} (\bibinfo {year} {2009})}\BibitemShut {NoStop}%
\bibitem [{\citenamefont {Slavin}\ and\ \citenamefont
  {Tiberkevich}(2009)}]{SlavinTutorial}%
  \BibitemOpen
  \bibfield  {author} {\bibinfo {author} {\bibfnamefont {A.}~\bibnamefont
  {Slavin}}\ and\ \bibinfo {author} {\bibfnamefont {V.}~\bibnamefont
  {Tiberkevich}},\ }
  {\bibfield  {journal} {\bibinfo  {journal}
  {IEEE Trans. Magn.},\ }\textbf {\bibinfo {volume} {45}},\ \bibinfo {pages}
  {1875} (\bibinfo {year} {2009})}\BibitemShut {NoStop}%
\bibitem [{\citenamefont {Consolo}\ \emph {et~al.}(2010)\citenamefont
  {Consolo}, \citenamefont {Puliafito}, \citenamefont {Finocchio},
  \citenamefont {Lopez~Diaz}, \citenamefont {Zivieri}, \citenamefont
  {Giovannini}, \citenamefont {Nizzoli}, \citenamefont {Valenti},\ and\
  \citenamefont {Azzerboni}}]{consolo2010ieee}%
  \BibitemOpen
  \bibfield  {author} {\bibinfo {author} {\bibfnamefont {G.}~\bibnamefont
  {Consolo}}, \bibinfo {author} {\bibfnamefont {V.}~\bibnamefont {Puliafito}},
  \bibinfo {author} {\bibfnamefont {G.}~\bibnamefont {Finocchio}}, \bibinfo
  {author} {\bibfnamefont {L.}~\bibnamefont {Lopez~Diaz}}, \bibinfo {author}
  {\bibfnamefont {R.}~\bibnamefont {Zivieri}}, \bibinfo {author} {\bibfnamefont
  {L.}~\bibnamefont {Giovannini}}, \bibinfo {author} {\bibfnamefont
  {F.}~\bibnamefont {Nizzoli}}, \bibinfo {author} {\bibfnamefont
  {G.}~\bibnamefont {Valenti}}, \ and\ \bibinfo {author} {\bibfnamefont
  {B.}~\bibnamefont {Azzerboni}},\ }
  {\bibfield  {journal} {\bibinfo  {journal} {IEEE Trans. Magn.},\ }\textbf
  {\bibinfo {volume} {46}},\ \bibinfo {pages} {3629} (\bibinfo {year}
  {2010})}\BibitemShut {NoStop}%
\bibitem [{\citenamefont {Pufall}\ \emph {et~al.}(2005)\citenamefont {Pufall},
  \citenamefont {Rippard}, \citenamefont {Kaka}, \citenamefont {Silva},\ and\
  \citenamefont {Russek}}]{Pufall2005APL}%
  \BibitemOpen
  \bibfield  {author} {\bibinfo {author} {\bibfnamefont {M.~R.}\ \bibnamefont
  {Pufall}}, \bibinfo {author} {\bibfnamefont {W.~H.}\ \bibnamefont {Rippard}},
  \bibinfo {author} {\bibfnamefont {S.}~\bibnamefont {Kaka}}, \bibinfo {author}
  {\bibfnamefont {T.~J.}\ \bibnamefont {Silva}}, \ and\ \bibinfo {author}
  {\bibfnamefont {S.~E.}\ \bibnamefont {Russek}},\ }
  {\bibfield  {journal} {\bibinfo
  {journal} {Appl. Phys. Lett.},\ }\textbf {\bibinfo {volume} {86}},\ \bibinfo
  {pages} {082506} (\bibinfo {year} {2005})}\BibitemShut {NoStop}%
\bibitem [{\citenamefont {Muduli}\ \emph {et~al.}(2010)\citenamefont {Muduli},
  \citenamefont {Pogoryelov}, \citenamefont {Bonetti}, \citenamefont {Consolo},
  \citenamefont {Mancoff},\ and\ \citenamefont {\AA{}kerman}}]{muduli2010prb}%
  \BibitemOpen
  \bibfield  {author} {\bibinfo {author} {\bibfnamefont {P.~K.}\ \bibnamefont
  {Muduli}}, \bibinfo {author} {\bibfnamefont {Y.}~\bibnamefont {Pogoryelov}},
  \bibinfo {author} {\bibfnamefont {S.}~\bibnamefont {Bonetti}}, \bibinfo
  {author} {\bibfnamefont {G.}~\bibnamefont {Consolo}}, \bibinfo {author}
  {\bibfnamefont {F.}~\bibnamefont {Mancoff}}, \ and\ \bibinfo {author}
  {\bibfnamefont {J.}~\bibnamefont {\AA{}kerman}},\ }
  {\bibfield  {journal} {\bibinfo  {journal}
  {Phys. Rev. B},\ }\textbf {\bibinfo {volume} {81}},\ \bibinfo {pages}
  {140408(R)} (\bibinfo {year} {2010})}\BibitemShut {NoStop}%
\bibitem [{\citenamefont {Slavin}\ and\ \citenamefont
  {Tiberkevich}(2005){\natexlab{a}}}]{Slavin2005aPRB}%
  \BibitemOpen
  \bibfield  {author} {\bibinfo {author} {\bibfnamefont {A.~N.}\ \bibnamefont
  {Slavin}}\ and\ \bibinfo {author} {\bibfnamefont {V.~S.}\ \bibnamefont
  {Tiberkevich}},\ }
  {\bibfield  {journal}
  {\bibinfo  {journal} {Phys. Rev. B},\ }\textbf {\bibinfo {volume} {72}},\
  \bibinfo {pages} {092407} (\bibinfo {year} {2005}{\natexlab{a}})}\BibitemShut
  {NoStop}%
\bibitem [{\citenamefont {Grollier}\ \emph {et~al.}(2006)\citenamefont
  {Grollier}, \citenamefont {Cros},\ and\ \citenamefont
  {Fert}}]{grollierj2006}%
  \BibitemOpen
  \bibfield  {author} {\bibinfo {author} {\bibfnamefont {J.}~\bibnamefont
  {Grollier}}, \bibinfo {author} {\bibfnamefont {V.}~\bibnamefont {Cros}}, \
  and\ \bibinfo {author} {\bibfnamefont {A.}~\bibnamefont {Fert}},\ }
  {\bibfield  {journal} {\bibinfo  {journal}
  {Phys. Rev. B},\ }\textbf {\bibinfo {volume} {73}},\ \bibinfo {pages}
  {060409(R)} (\bibinfo {year} {2006})}\BibitemShut {NoStop}%
\bibitem [{\citenamefont {Tiberkevich}\ \emph {et~al.}(2009)\citenamefont
  {Tiberkevich}, \citenamefont {Slavin}, \citenamefont {Bankowski},\ and\
  \citenamefont {Gerhart}}]{Tiberkevich2009apl}%
  \BibitemOpen
  \bibfield  {author} {\bibinfo {author} {\bibfnamefont {V.}~\bibnamefont
  {Tiberkevich}}, \bibinfo {author} {\bibfnamefont {A.}~\bibnamefont {Slavin}},
  \bibinfo {author} {\bibfnamefont {E.}~\bibnamefont {Bankowski}}, \ and\
  \bibinfo {author} {\bibfnamefont {G.}~\bibnamefont {Gerhart}},\ }
  {\bibfield  {journal} {\bibinfo  {journal} {Appl. Phys. Lett.},\ }\textbf
  {\bibinfo {volume} {95}},\ \bibinfo {eid} {262505} (\bibinfo {year}
  {2009})}\BibitemShut {NoStop}%
\bibitem [{\citenamefont {Bonin}\ \emph {et~al.}(2009)\citenamefont {Bonin},
  \citenamefont {Bertotti}, \citenamefont {Serpico}, \citenamefont
  {Mayergoyz},\ and\ \citenamefont {D'Aquino}}]{Bonin2009EPJB}%
  \BibitemOpen
  \bibfield  {author} {\bibinfo {author} {\bibfnamefont {R.}~\bibnamefont
  {Bonin}}, \bibinfo {author} {\bibfnamefont {G.}~\bibnamefont {Bertotti}},
  \bibinfo {author} {\bibfnamefont {C.}~\bibnamefont {Serpico}}, \bibinfo
  {author} {\bibfnamefont {I.~D.}\ \bibnamefont {Mayergoyz}}, \ and\ \bibinfo
  {author} {\bibfnamefont {M.}~\bibnamefont {D'Aquino}},\ }
  {\bibfield  {journal} {\bibinfo  {journal} {Eur.
  Phys. J. B},\ }\textbf {\bibinfo {volume} {68}},\ \bibinfo {pages} {221}
  (\bibinfo {year} {2009})}\BibitemShut {NoStop}%
\bibitem [{\citenamefont {Sano}(2007)}]{Sano2007jjap}%
  \BibitemOpen
  \bibfield  {author} {\bibinfo {author} {\bibfnamefont {E.}~\bibnamefont
  {Sano}},\ }
  {\bibfield  {journal} {\bibinfo  {journal}
  {Jpn. J. Appl. Phys.},\ }\textbf {\bibinfo {volume} {46}},\ \bibinfo {pages}
  {L1123} (\bibinfo {year} {2007})}\BibitemShut {NoStop}%
\bibitem [{\citenamefont {Georges}\ \emph {et~al.}(2008)\citenamefont
  {Georges}, \citenamefont {Grollier}, \citenamefont {Cros},\ and\
  \citenamefont {Fert}}]{Georges2008}%
  \BibitemOpen
  \bibfield  {author} {\bibinfo {author} {\bibfnamefont {B.}~\bibnamefont
  {Georges}}, \bibinfo {author} {\bibfnamefont {J.}~\bibnamefont {Grollier}},
  \bibinfo {author} {\bibfnamefont {V.}~\bibnamefont {Cros}}, \ and\ \bibinfo
  {author} {\bibfnamefont {A.}~\bibnamefont {Fert}},\ }
  {\bibfield  {journal} {\bibinfo  {journal} {Appl.
  Phys. Lett.},\ }\textbf {\bibinfo {volume} {92}},\ \bibinfo {pages} {232504}
  (\bibinfo {year} {2008})}\BibitemShut {NoStop}%
\bibitem [{\citenamefont {Persson}\ \emph {et~al.}(2007)\citenamefont
  {Persson}, \citenamefont {Zhou},\ and\ \citenamefont
  {{\AA}kerman}}]{Persson2007}%
  \BibitemOpen
  \bibfield  {author} {\bibinfo {author} {\bibfnamefont {J.}~\bibnamefont
  {Persson}}, \bibinfo {author} {\bibfnamefont {Y.}~\bibnamefont {Zhou}}, \
  and\ \bibinfo {author} {\bibfnamefont {J.}~\bibnamefont {{\AA}kerman}},\
  }
  {\bibfield  {journal} {\bibinfo  {journal} {J. Appl.
  Phys.},\ }\textbf {\bibinfo {volume} {101}},\ \bibinfo {pages} {09A503}
  (\bibinfo {year} {2007})}\BibitemShut {NoStop}%
\bibitem [{\citenamefont {Zhou}\ \emph {et~al.}(2007)\citenamefont {Zhou},
  \citenamefont {Persson},\ and\ \citenamefont {{\AA}kerman}}]{Zhou2007}%
  \BibitemOpen
  \bibfield  {author} {\bibinfo {author} {\bibfnamefont {Y.}~\bibnamefont
  {Zhou}}, \bibinfo {author} {\bibfnamefont {J.}~\bibnamefont {Persson}}, \
  and\ \bibinfo {author} {\bibfnamefont {J.}~\bibnamefont {{\AA}kerman}},\
  }
  {\bibfield  {journal} {\bibinfo  {journal} {J. Appl.
  Phys.},\ }\textbf {\bibinfo {volume} {101}},\ \bibinfo {pages} {09A510}
  (\bibinfo {year} {2007})}\BibitemShut {NoStop}%
\bibitem [{\citenamefont {Zhou}\ \emph {et~al.}(2008)\citenamefont {Zhou},
  \citenamefont {Persson}, \citenamefont {Bonetti},\ and\ \citenamefont
  {{\AA}kerman}}]{Zhou2008}%
  \BibitemOpen
  \bibfield  {author} {\bibinfo {author} {\bibfnamefont {Y.}~\bibnamefont
  {Zhou}}, \bibinfo {author} {\bibfnamefont {J.}~\bibnamefont {Persson}},
  \bibinfo {author} {\bibfnamefont {S.}~\bibnamefont {Bonetti}}, \ and\
  \bibinfo {author} {\bibfnamefont {J.}~\bibnamefont {{\AA}kerman}},\
  }
  {\bibfield  {journal}
  {\bibinfo  {journal} {Appl. Phys. Lett.},\ }\textbf {\bibinfo {volume}
  {92}},\ \bibinfo {eid} {092505} (\bibinfo {year} {2008})}\BibitemShut
  {NoStop}%
\bibitem [{\citenamefont {Zhou}\ \emph {et~al.}(2009)\citenamefont {Zhou},
  \citenamefont {Bonetti}, \citenamefont {Persson},\ and\ \citenamefont
  {{\AA}kerman}}]{Zhou2009ieee}%
  \BibitemOpen
  \bibfield  {author} {\bibinfo {author} {\bibfnamefont {Y.}~\bibnamefont
  {Zhou}}, \bibinfo {author} {\bibfnamefont {S.}~\bibnamefont {Bonetti}},
  \bibinfo {author} {\bibfnamefont {J.}~\bibnamefont {Persson}}, \ and\
  \bibinfo {author} {\bibfnamefont {J.}~\bibnamefont {{\AA}kerman}},\
  }
  {\bibfield  {journal} {\bibinfo  {journal} {IEEE
  Trans. Magn.},\ }\textbf {\bibinfo {volume} {45}},\ \bibinfo {pages} {2421}
  (\bibinfo {year} {2009})}\BibitemShut
  {NoStop}%
\bibitem [{\citenamefont {Zhou}\ and\ \citenamefont
  {{\AA}kerman}(2009)}]{zhou2009apl}%
  \BibitemOpen
  \bibfield  {author} {\bibinfo {author} {\bibfnamefont {Y.}~\bibnamefont
  {Zhou}}\ and\ \bibinfo {author} {\bibfnamefont {J.}~\bibnamefont
  {{\AA}kerman}},\ }
  {\bibfield  {journal} {\bibinfo  {journal} {Appl. Phys. Lett.},\ }\textbf
  {\bibinfo {volume} {94}},\ \bibinfo {eid} {112503} (\bibinfo {year}
  {2009})}\BibitemShut {NoStop}%
\bibitem [{\citenamefont {Bonetti}\ \emph
  {et~al.}(2009){\natexlab{a}}\citenamefont {Bonetti}, \citenamefont {Muduli},
  \citenamefont {Mancoff},\ and\ \citenamefont {{\AA}kerman}}]{bonetti2009apl}%
  \BibitemOpen
  \bibfield  {author} {\bibinfo {author} {\bibfnamefont {S.}~\bibnamefont
  {Bonetti}}, \bibinfo {author} {\bibfnamefont {P.}~\bibnamefont {Muduli}},
  \bibinfo {author} {\bibfnamefont {F.}~\bibnamefont {Mancoff}}, \ and\
  \bibinfo {author} {\bibfnamefont {J.}~\bibnamefont {{\AA}kerman}},\
  }
  {\bibfield  {journal} {\bibinfo  {journal}
  {Appl. Phys. Lett.},\ }\textbf {\bibinfo {volume} {94}},\ \bibinfo {eid}
  {102507} (\bibinfo {year} {2009})}\BibitemShut {NoStop}%
\bibitem [{\citenamefont {Slonczewski}(1999)}]{Slonczewski1999}%
  \BibitemOpen
  \bibfield  {author} {\bibinfo {author} {\bibfnamefont {J.~C.}\ \bibnamefont
  {Slonczewski}},\ }
  {\bibfield  {journal} {\bibinfo  {journal}
  {J. Magn. Magn. Mater.},\ }\textbf {\bibinfo {volume} {195}},\ \bibinfo
  {pages} {261} (\bibinfo {year} {1999})}\BibitemShut {NoStop}%
\bibitem [{\citenamefont {Slavin}\ and\ \citenamefont
  {Tiberkevich}(2005){\natexlab{b}}}]{Slavin2005PRL}%
  \BibitemOpen
  \bibfield  {author} {\bibinfo {author} {\bibfnamefont {A.}~\bibnamefont
  {Slavin}}\ and\ \bibinfo {author} {\bibfnamefont {V.}~\bibnamefont
  {Tiberkevich}},\ }
  {\bibfield  {journal} {\bibinfo  {journal} {Phys. Rev. Lett.},\ }\textbf
  {\bibinfo {volume} {95}},\ \bibinfo {pages} {237201} (\bibinfo {year}
  {2005})}\BibitemShut {NoStop}%
\bibitem [{\citenamefont {Bonetti}\ \emph {et~al.}(2010)\citenamefont
  {Bonetti}, \citenamefont {Tiberkevich}, \citenamefont {Consolo},
  \citenamefont {Finocchio}, \citenamefont {Muduli}, \citenamefont {Mancoff},
  \citenamefont {Slavin},\ and\ \citenamefont {\AA~kerman}}]{Bonetti2010prl}%
  \BibitemOpen
  \bibfield  {author} {\bibinfo {author} {\bibfnamefont {S.}~\bibnamefont
  {Bonetti}}, \bibinfo {author} {\bibfnamefont {V.}~\bibnamefont
  {Tiberkevich}}, \bibinfo {author} {\bibfnamefont {G.}~\bibnamefont
  {Consolo}}, \bibinfo {author} {\bibfnamefont {G.}~\bibnamefont {Finocchio}},
  \bibinfo {author} {\bibfnamefont {P.}~\bibnamefont {Muduli}}, \bibinfo
  {author} {\bibfnamefont {F.}~\bibnamefont {Mancoff}}, \bibinfo {author}
  {\bibfnamefont {A.}~\bibnamefont {Slavin}}, \ and\ \bibinfo {author}
  {\bibfnamefont {J.}~\bibnamefont {\AA~kerman}},\ }
  {\bibfield  {journal}
  {\bibinfo  {journal} {Phys. Rev. Lett.},\ }\textbf {\bibinfo {volume}
  {105}},\ \bibinfo {pages} {217204} (\bibinfo {year} {2010})}\BibitemShut {NoStop}%
\bibitem [{\citenamefont {Russek}\ \emph {et~al.}(2005)\citenamefont {Russek},
  \citenamefont {Kaka}, \citenamefont {Rippard}, \citenamefont {Pufall},\ and\
  \citenamefont {Silva}}]{Russek2005PRB}%
  \BibitemOpen
  \bibfield  {author} {\bibinfo {author} {\bibfnamefont {S.~E.}\ \bibnamefont
  {Russek}}, \bibinfo {author} {\bibfnamefont {S.}~\bibnamefont {Kaka}},
  \bibinfo {author} {\bibfnamefont {W.~H.}\ \bibnamefont {Rippard}}, \bibinfo
  {author} {\bibfnamefont {M.~R.}\ \bibnamefont {Pufall}}, \ and\ \bibinfo
  {author} {\bibfnamefont {T.~J.}\ \bibnamefont {Silva}},\ }
  {\bibfield
  {journal} {\bibinfo  {journal} {Phys. Rev. B},\ }\textbf {\bibinfo {volume}
  {71}},\ \bibinfo {pages} {104425} (\bibinfo {year} {2005})}\BibitemShut
  {NoStop}%
\bibitem [{\citenamefont {Silva}\ and\ \citenamefont
  {Keller}(2010)}]{Silva2010ieee}%
  \BibitemOpen
  \bibfield  {author} {\bibinfo {author} {\bibfnamefont {J.}~\bibnamefont
  {Silva}}\ and\ \bibinfo {author} {\bibfnamefont {W.}~\bibnamefont {Keller}},\
  }
  {\bibfield  {journal} {\bibinfo  {journal} {IEEE
  Trans. Magn.},\ }\textbf {\bibinfo {volume} {46}},\ \bibinfo {pages} {3555}
  (\bibinfo {year} {2010})}\BibitemShut
  {NoStop}%
\bibitem [{\citenamefont {Pikovsky2001}\ \emph {et~al.}(2001)\citenamefont
  {Pikovsky}, \citenamefont {Rosenblum},\ and\ \citenamefont
  {Kurths}}]{Pikovsky2001}%
  \BibitemOpen
  \bibfield  {author} {\bibinfo {author} {\bibfnamefont {A.}~\bibnamefont
  {Pikovsky}}, \bibinfo {author} {\bibfnamefont {M.}~\bibnamefont {Rosenblum}},
  \ and\ \bibinfo {author} {\bibfnamefont {J.}~\bibnamefont {Kurths}},\ }
  {\emph   {\bibinfo {title} {Synchronization. A universal concept in nonlinear
  sciences}}},\ edited by\ \bibinfo {editor} {\bibfnamefont {B.}~\bibnamefont
  {Chirikov}}, \bibinfo {editor} {\bibfnamefont {P.}~\bibnamefont
  {Cvitanovic}}, \bibinfo {editor} {\bibfnamefont {F.}~\bibnamefont {Moss}}, \
  and\ \bibinfo {editor} {\bibfnamefont {H.}~\bibnamefont {Swinney}},
  \ (\bibinfo  {publisher}
  {Cambridge University Press, Cambridge},\ \bibinfo {year} {2001})\ 
  \BibitemShut {NoStop}%
\end{thebibliography}

%

\end{document}